# An ionization chamber for (n,z) reaction cross section measurements on gaseous targets

R. Machrafi, *Guinyun Kim, Dongchul Son*
*Center of High Energy Physics Kyunpook National University, 1370, Sankyuk-Dong, Pukgu*
*Taegu, S.Korea*
*Yu.M.Gledenov, V.I.Salatski, P.V.Sedyshev*
*JINR Dubna, 980141 Moscow region, Russia*
*J.Andrzejewski, P.J.Szalanski*
*University of Lodz, 90131 Lodz, Poland*

**Abstract**

An ionization chamber with gaseous samples has been designed. It has been tested on the beam of the pulsed reactor IBR-30 of FLNP, JINR-Dubna. The experiment has been carried out with resonance neutrons. The exposed gas volume serves as a target for neutron beam. We have compared the chamber to samples on substrates, the background component due to Li and B microimpurities in this case is totally absent. It has been tested also the recovery capability of the chamber after the reactor power pulse using the protons from the $^{3}He(n,p)^{3}H$ reaction, α-particles from a U-source and a pulsed precision generator. Moreover the energy resolution of the chamber with its equipment has been carried out.

**Introduction**

Lately it has been designed an ionization chamber with a grid (GIC) for gaseous target studies. On the basis of this chamber at the IBR-2 pulsed reactor [2], it has been measured the thermal cross section of $^{17}O(n,\alpha)^{14}C$, $^{36}Ar(n,\alpha)^{33}S$ and $^{21}Ne(n,\alpha)^{18}O$ reactions [1]. The method has been very effective for studies of neutron reactions with emission of charged particles. To complete the methodical studies i.e. to investigate the range of higher energies, we have tested the constructed chamber with resonance neutrons of the IBR-30 pulsed reactor, FLNP, JINR-Dubna. The gas of the chamber serves as a gaseous target for neutron beam. For thermal neutron, we have compared the GIC with that using solid samples on the substrate

This article discuses the characteristics of the chamber carried out with resonance neutrons, mainly : the energy resolution, its setup time after the power pulse of the reactor and its comparison with that using solid samples for thermal neutrons.

**Experiment**

To determine the characteristics of the constructed chamber in measurements with resonance neutrons, we have used the neutrons from the IBR-30 pulsed reactor operating with a linear electron accelerator. The chamber was installed at a distance of 30 m from the reactor (Fig.1.) and it was filled with a mixture of 93% Ar, 5% $CO_2$ and 2% $^{3}He$ at a pressure of 1.8 ata. To reduce the background, 12.5 cm thick Al filter was placed 10 m apart from the reactor and in front of the chamber there was put 2 mm thick Cd filter. The time-of-flight calibration of the neutron energies was done using Mn and In filters placed in front of the chamber. In the thermal neutron measurements, to compare the GIC spectra with those obtained using the ionization chamber with a target on a substrate, the Cd filter was removed. The neutron beam transmitted through the chamber was formed using 20x40 mm collimators in front of the chamber. Inside the chamber an α-particle uranium source ($^{238}$U and $^{234}$U mixture) was fixed and a precision pulse generator was connected to the test input of the collector preamplifier. Having used a multiparameter module [3], we have registered in coincidence amplitude signals of the protons from the $^3$He(n,p)$^3$H reaction, of the α-particles from the uranium source, the pulses from a precision pulse generator and corresponding time-of-flight signals. Figure.2 shows the obtained pulse height spectra. By setting a window on the amplitude spectrum one can get the corresponding yield as a function of neutron energy. The capability of the chamber to recover its normal working mode was tested after the power pulse of the reactor, during which a powerful flux of γ-quanta and fast neutrons creates an overload on the chamber.

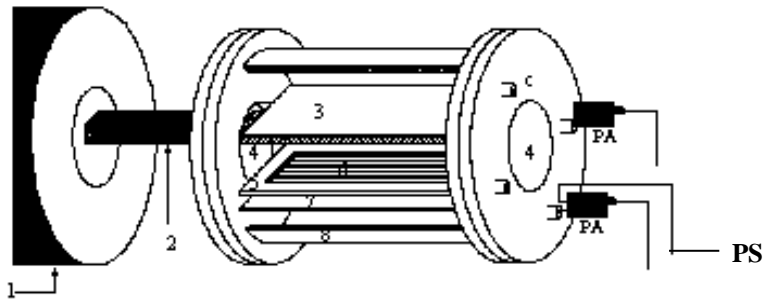

**Fig.1.** The experimental setup of the measurement: 1- neutron guide, 2– neutron beam, 3– cathode, 4- inlet and outlet windows, 5– grid frame, 6- grid, 7- collector (anode), 8- stainless, c- connector, PA- preamplifier, A-amplifier, PS-pulse shaper

Figure.3 illustrates the time-of-flight spectrum for an amplitude spectrum window of α-particles from the uranium source. It is seen that, in 20 μs after the reactor power pulse, the chamber restores the normal operation mode. To increase the number of counts in each channel of the time-of-flight spectrum (1μs channel wide), we used a precision pulse generator with the necessary pulses frequency. The obtained spectrum is also shown in Fig.3. It helps to determine more exactly the recovery time of the chamber together with the used equipment. This time appeared to be about 26 μs.

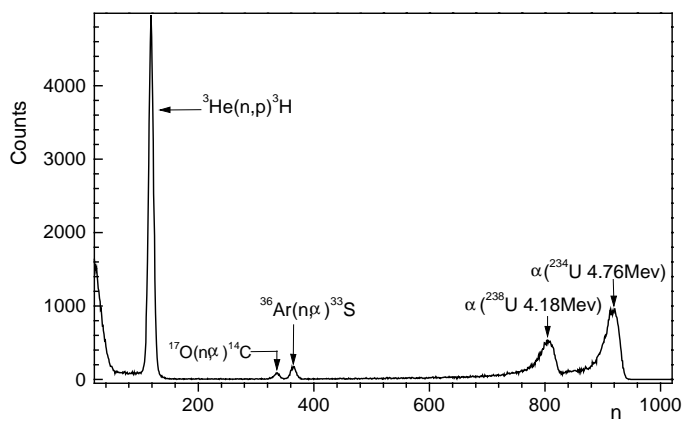

**Fig.2**. Pulse height spectra in the energy range up to ~1MeV

The dependence of the resolution of the peaks from the $^3$He(n,p)$^3$H reaction products on the neutron energy was also determined (Fig.4).

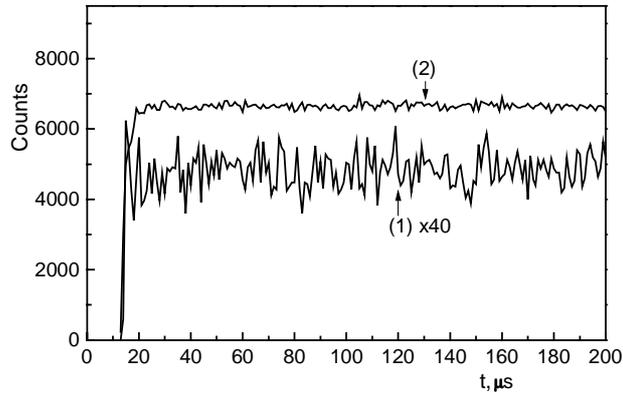

**Fig.3**. Time-of-flight spectra, (1)- α-particles from the uranium source; (2)- pulses from the pulser; t - is the neutron time of flight

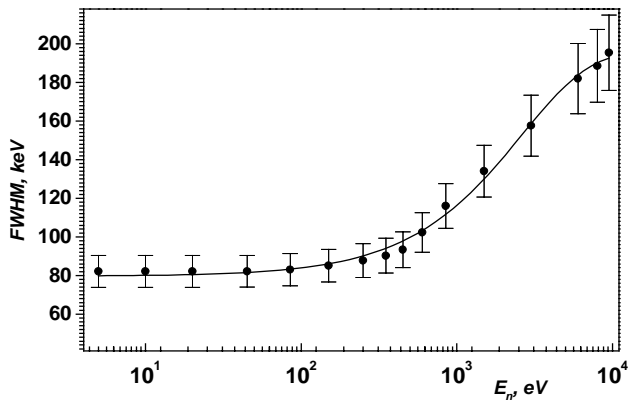

**Fig.4.** The energy resolution of peaks from the $^3$He(n,p)$^3$H reaction products as a function of the neutron energy.

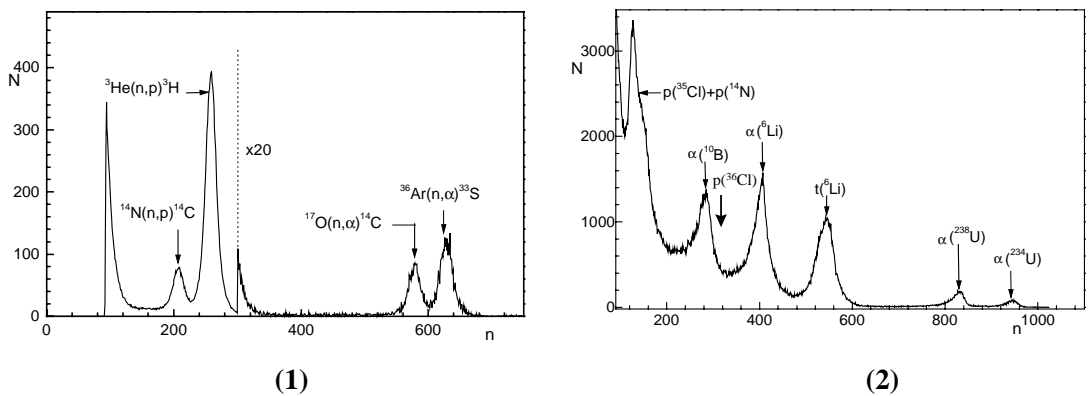

(1)                                                                 (2)

**Puc.5.** The comparison of the pulse-height spectra carried out with thermal neutrons.
 (1) the chamber using gaseous targets: 93.75% Ar, 6.25% $CO_2$ and $1.8 \cdot 10^{-5}$% $^3$He
 (2) the chamber using solid target: NaCl solid target (enriched by $^{36}$Cl)

**Discussion of the results**

The FWHM for the IBR-30 reactor pulse is 4.5 μs[4], while at 10% of the height, it is 20 μs because of the back front delay. It is seen that, approximately, for the neutron energy up to 1 keV, the resolution keeps the level of 80-100 keV (Fig.3). However, approaching the reactor neutron pulse, it increases rather fast to about 200 keV at $E_n \approx 10$ keV. Before, On the same 30m flight path, the experiments carried out using the chamber with solid samples, did not allow researchers to reach an energy of 1 keV with this resolution. We have compared the spectra of the particles from the reaction obtained in thermal neutron measurements using GIC and those when the used chamber was with a solid target on a substrate. Figure.5. shows the spectrum obtained with the ionization chamber containing NaCl solid target (enriched by $^{36}$Cl) and an α-uranium source. It shows peaks from the extremely small $^{10}$B and $^{6}$Li impurities that mask the peak of the protons from the reaction $^{36}$Cl(n,p)$^{36}$S. Such peaks were not observed in the case of GIC at a gas pressure of 1.15 ata and higher [1]. As it has been stated, in the test measurements, as the pressure decreased to 0.6 ata, there appeared peaks in the spectra due to the tracks of the tritons (from the reaction on $^{6}$Li in the aluminum windows) which had reached the sensitive volume [1].

**Conclusion**

In comparison with an ionization chamber with solid target (see for example the difficulties meeting in Ref.[5] to prepare the target in the measurement of $^{17}$O(n,α)$^{14}$C cross-section) our method using the gaseous targets has the following advantages:
- lower background;
- no peaks from the reactions on the $^{10}$B and $^{6}$Li impurities from support material (substrate)
- essentially simpler, more exact and reliable determination of the number of nuclei in the investigated sample;
- no tails in the region of lower energies in the spectrum of charged particles (Fig.2) because of the thickness of the target.
- charged particles are registered in 4π-geometry (very large geometrical efficiency close to 100%).

On the 30 m flight path of the reactor IBR-30, even for about 20 μs after the pulse of the reactor, i.e. for neutron energy up to 10 keV, the energy resolution is still lower than 200 keV (60-80 keV without beam). For investigation of many reactions, this is quite sufficient. If the need arises to have a better resolution or investigate the range of higher energies, it is necessary to use longer flight paths or neutron sources with shorter pulses. The method will be adopted for getting neutron data at Pohang neutron facility in Korea.